# Influence of composition and precipitation evolution on damage at grain boundaries in a crept polycrystalline Ni-based superalloy

Paraskevas Kontis[a,*], Aleksander Kostka[a,b], Dierk Raabe[a], Baptiste Gault[a],

[a] Max-Planck-Institut für Eisenforschung, Max-Planck-Str. 1, 40237 Düsseldorf, Germany.

[b] Materials Research Department and Centre for Interface Dominated Materials (ZGH), Ruhr-University Bochum, 44801, Bochum, Germany

**Abstract**

The microstructural and compositional evolution of intergranular carbides and borides prior to and after creep deformation at 850 °C in a polycrystalline nickel-based superalloy was studied. Primary MC carbides, enveloped within intergranular γ' layers, decomposed resulting in the formation of layers of the undesirable η phase. These layers have a composition corresponding to $Ni_3Ta$ as measured by atom probe tomography and their structure is consistent with the $D0_{24}$ hexagonal structure as revealed by transmission electron microscopy. Electron backscattered diffraction reveals that they assume various misorientations with regard to the adjacent grains. As a consequence, these layers act as brittle recrystallized zones and crack initiation sites. The composition of the MC carbides after creep was altered substantially, with the Ta content decreasing and the Hf and Zr contents increasing, suggesting a beneficial effect of Hf and Zr additions on the stability of MC carbides. By contrast, $M_5B_3$ borides were found to be microstructurally stable after creep and without substantial compositional changes. Borides at 850 °C were found to coarsen, resulting in some cases into γ'- depleted zones, where, however, no cracks were observed. The major consequences of secondary phases on the microstructural stability of superalloys during the design of new polycrystalline superalloys are discussed.

[*] Corresponding author: **p.kontis@mpie.de**

# 1. Introduction

Superalloys are used for multiple critical components in aero-engines and land-based turbines operated over prolonged exposure times at elevated temperatures. Maintaining the microstructural integrity of these materials during operation is a paramount challenge in this field [1,2]. Most investigations of the microstructural changes during high temperature deformation focus on the main precipitation strengthening phase in the superalloys, the γ' precipitates, and its evolution over time in terms of size [3,4], volume fraction [5,6], shape [7,8], composition [9–11] and its different deformation mechanisms occurring under various thermal loading scenarios [12–16]. However, little attention has been paid so far to the minor elements, such as carbon and boron and their associated phases, which can also have major consequences for the mechanical performance of superalloys. If not correctly predicted and controlled, the formation and precipitation of "secondary" phases, such as carbides and borides for instance along grain boundaries, can result in premature fracture of superalloys, occurring prior to deformation of γ' precipitates [17]. We therefore aim here at obtaining a better fundamental understanding of the microstructural and compositional evolution of these secondary phases during deformation at high temperature, which will allow us to design superalloys withstanding higher operational temperatures and longer exposure times.

Traditionally, in polycrystalline superalloys, two grain boundary strengthening elements are added. First, carbon is added and usually leads to the formation of carbides. Carbides are used to control grain size during thermomechanical processing [18,19] and to prevent grain boundary sliding during deformation [20,21]. However, there are many cases in which carbides are detrimental to the performance of superalloys by acting as stress concentration sites resulting in the initiation of cracks [22,23]. Besides, rapid oxidation of MC carbides at elevated temperatures leads to the formation of porous, brittle oxides and soft recrystallized grains in their vicinity [24–26]. Decomposition of MC carbides often follows protracted periods of service exposure, resulting in the formation of either different types of carbides [27,28] or topologically closed-packed (TCP) phases [29–31]. In particular, the formation of TCP phases such as eta (η) is often considered detrimental to the mechanical performance of superalloys [32,33]. Second, boron is employed as a grain boundary strengthening element in polycrystalline superalloys. Depending on the grain size and bulk composition of the alloy, boron can be found either segregating at grain boundaries [34,35] or participating to the formation of various types of borides such as $M_5B_3$ and $M_3B_2$ along the grain boundaries [36–38]. Borides are considered as secondary phases in superalloys, but their presence or absence in the microstructure can control the mechanical performance of polycrystalline superalloys [39,40]. Although carbides and borides are known to significantly influence the mechanical response of superalloys, there is only very little information about their chemical and microstructural evolution during prolonged deformation times at elevated temperatures.

In this study, we have systematically investigated the microstructural and compositional evolution of precipitates formed at grain boundaries in a polycrystalline nickel-based superalloy in the heat-treated condition and after creep deformation at 850 °C for approximately 3000 h. We have observed the decomposition of intergranular Ta-rich MC carbides resulting in layers of the TCP η phase along the grain boundaries. Electron-backscattered diffraction analysis revealed various orientations for these layers with respect to the adjacent grains, acting as recrystallized regions consisting of the brittle η phase along the grain boundaries. The composition of the MC carbides has substantially evolved after their decomposition as confirmed by atom probe tomography (APT). In contrast, borides were found to be more stable compared to MC carbides, retaining their primary composition. The ramifications of the observed microstructural and chemical grain boundary evolutions on the creep performance are discussed.

## 2. Experimental procedures
### 2.1. Materials and mechanical testing
A polycrystalline nickel-based superalloy STAL15-CC was studied with a composition of Ni-16.50Cr-5.50Co-0.60Mo-1.20W-10.00Al-2.40Ta-0.02Hf-0.5C-0.05B-0.01Zr (at.%) and a grain size approximately 750 μm on average. A hot isostatic press step (HIP) at 1195 °C for 5 hours under 175 MPa pressure was initially performed to cast bars. A primary ageing at 1120 °C for 4 hours followed the HIP stage, and, subsequently, a second aging was performed at 845 °C for 24 hours, both followed by air cooling. Creep specimens with a gauge length of 25 mm were machined from the fully heat-treated bars and tested under constant load of 235MPa at 850 °C up to fracture, with strain at fracture recorded to be 2.25%. All the observations related to the crept sample were made at a distance of 1 mm from the fracture surface and within the gauge section.

## 2.2. High-resolution characterisation

The fully heat treated and crept microstructure were primarily investigated using a Zeiss Merlin scanning electron microscope (SEM), operated at 20kV in backscatter diffraction mode. In addition, electron backscattered diffraction (EBSD) analysis across grain boundaries in the crept alloy was performed in a Zeiss 1540 XB SEM. Specimens were polished to 0.4 µm colloidal silica finish and EBSD was performed at 20kV for which the exposure time was 0.5 s, with the scintillator screen held at a capture angle of 70° to the sample surface. Site-specific samples for transmission electron microscope (TEM) were prepared using focused ion beam (FIB) system FEI Helios G4 CX operated at 30kV. In the final specimen preparation step, low voltage (5kV) ion beam cleaning was applied for 2 minutes to each side of the TEM sample in order to remove regions severely affected by the ion beam damage. Microstructural analyses were performed on heat-treated and crept specimens using a FEI Tecnai Supertwin F20 TEM operated at 200kV. Finally, site specific lift-outs for atom probe tomography were prepared using a focused-ion beam FEI Helios 600 following procedures described in reference [41]. APT specimens were prepared from grain boundaries and within grains from both heat treated and creep conditions. The APT specimens were analysed in a Cameca 5000 XR instrument operated in laser pulsing mode at 60K, with a repetition rate varying between 125-200 kHz and with laser energy from 45 to 70 pJ, depending on the phase analysed in each case. Data reconstruction and processing was performed using the Cameca IVAS 3.8.2 software tool.

## 3. Results

### 3.1. Primary grain boundary phases after heat treatment

Figure 1a shows a backscattered SEM micrograph from a grain boundary of STAL15-CC after full heat treatment. A strong contrast based on the average atomic number Z between the different observed phases is readily apparent. The γ' precipitates within the grains can be clearly seen, while γ' layers along the grain boundary can also be distinguished. Besides the γ' precipitates, blocky particles with bright contrast and precipitates with dark contrast were also observed. These correspond to primary MC carbides and $M_5B_3$ borides as they were identified by TEM and APT, respectively. Note that MC carbides were also observed within the grains exhibiting both a script-like and blocky shape, as shown in Figure 1b. It is important to mention here that the intergranular MC carbides were mostly enveloped within γ' layers. This particular order at the grain boundaries will have substantial impact on the undesirable formation of the brittle η phase, as discussed below.

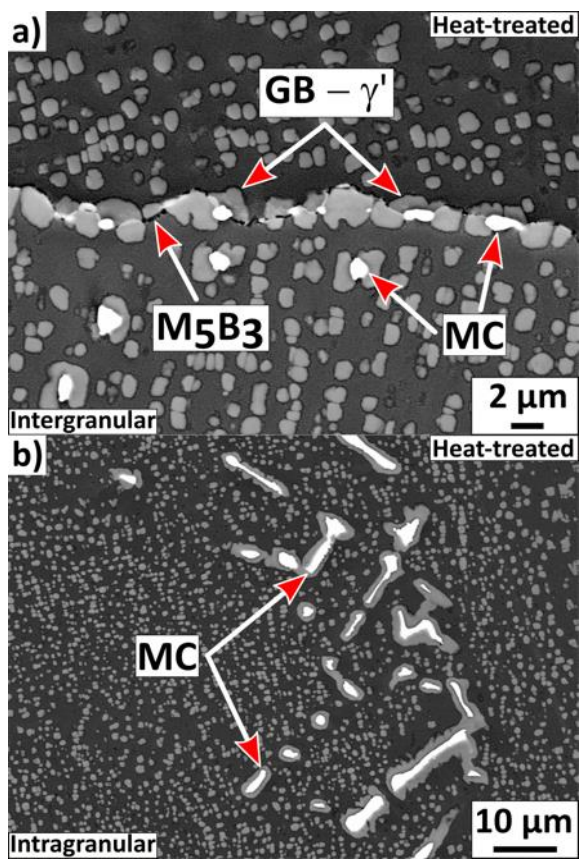

Figure 1: Backscattered electron micrographs of STAL15-CC alloy after heat treatment showing: a) the primary MC carbides and $M_5B_3$ borides along a grain boundary and b) primary MC-type carbides within a grain.

TEM analysis of the grain boundary precipitates after full heat treatment revealed precipitation of MC carbide with an FCC crystal structure, as shown in Figure 2. Figure 3a shows an APT 3D-reconstruction from an intergranular MC carbide, containing an interface between MC carbide and γ'. The corresponding composition profile across this interface revealed no segregation at this particular interface, as shown in Figure 3b. Thorough APT analyses of such precipitates has shown that the MC carbides are Ta-rich, yet, containing also other elements such as Hf, Zr, W and Mo participating in amounts lower than 1.7 at.%, as summarized in Table 1. As also seen from Table 1, the composition of MC carbides in three different datasets from different grain boundary particles varies slightly between particles. Small amounts of boron and nitrogen were also found within the primary MC. High amounts of nitrogen were observed by APT recently in MC carbides in a powder-processed nickel-based superalloy [42].

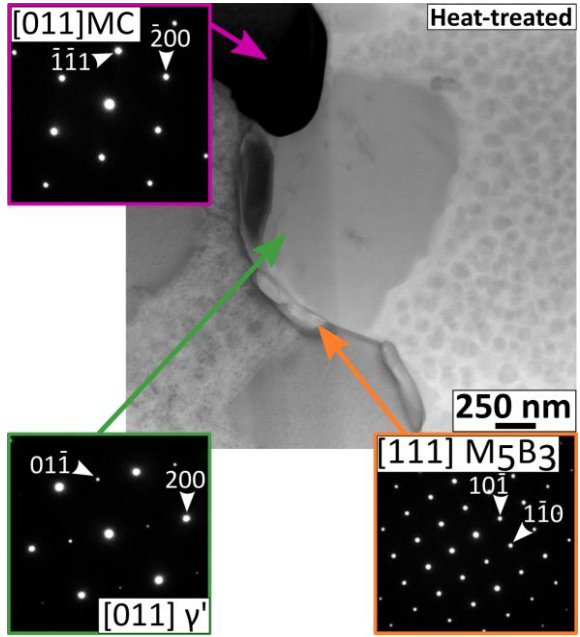

Figure 2: Transmission electron micrograph from a grain boundary with corresponding selected area diffraction patterns of MC, $M_5B_3$ and γ' precipitates after full heat treatment.

Besides the MC carbides, TEM investigations confirmed also the precipitation of intergranular $M_5B_3$ borides with a tetragonal structure as shown in Figure 2. APT analysis revealed that these borides are mainly Cr-rich whereas W and Mo contribute in lower amounts, as shown in Table 1. Although the observed borides consistently exhibit their typical tetragonal structure, their composition can vary, with the amounts of Cr, W and Mo showing significant variations from one precipitate to another. In particular, Cr varied between 52.0 and 58.0 at.%, W between 2.0 and 6.2 at.% and Mo between 2.4 and 4.3 at.%. A similar observation was previously reported regarding the composition of these secondary precipitates from an APT analysis of a powder nickel-based superalloy [38]. Note, that no $M_{23}C_6$ were observed along the grain boundaries of STAL15-CC in the full heat treated state, and that is also supported by a previous study [17].

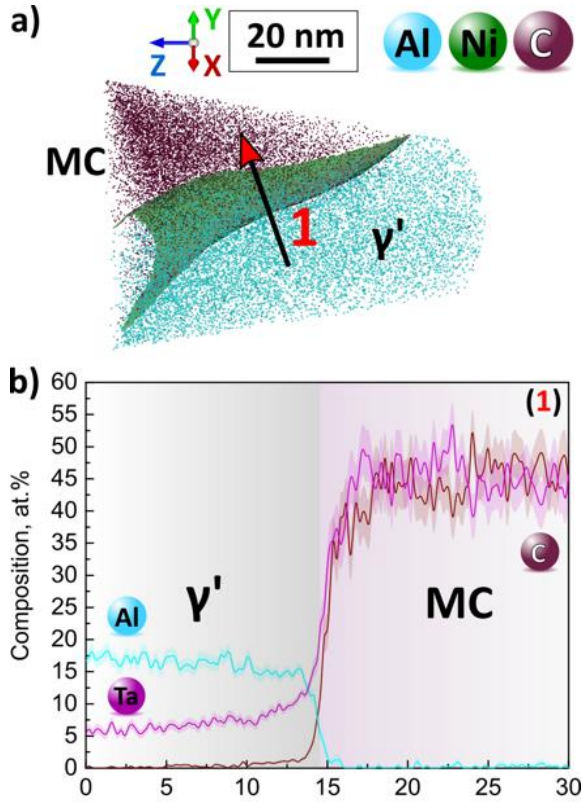

Figure 3: a) APT reconstruction from a grain boundary of STAL15-CC after full heat treatment, showing an MC/γ' interface with an isosurface at 57.0 at.% Ni. b) Composition profile across the MC/γ' interface corresponding to arrow #1 in Figure 3a. Error bars are shown as lines filled with colour and correspond to the 2σ counting error.

Table 1: Summary of composition of MC carbides after heat treatment from grain boundaries (GB) and within grains (GR) and intergranular $M_5B_3$ borides as collected by APT (at.%).

|  | C | Ta | Hf | Zr | W | Mo | Cr | B | N |
|---|---|---|---|---|---|---|---|---|---|
| MC-GB | 48.3±0.073 | 44.1±0.073 | 1.7±0.019 | 1.3±0.016 | 1.0±0.014 | 0.9±0.013 | 2.0±0.021 | 0.2±0.007 | 0.3±0.008 |
| MC-GB | 48.4±0.035 | 44.4±0.035 | 1.7±0.009 | 1.2±0.008 | 0.8±0.006 | 0.8±0.006 | 2.0±0.010 | 0.2±0.003 | 0.3±0.004 |
| MC-GB | 47.5±0.126 | 45.2±0.126 | 1.2±0.028 | 0.6±0.020 | 1.0±0.025 | 1.3±0.029 | 2.2±0.037 | 0.2±0.012 | 0.5±0.018 |
| MC-GR | 48.6±0.053 | 44.2±0.052 | 1.1±0.011 | 0.7±0.009 | 1.0±0.010 | 1.0±0.011 | 2.2±0.015 | 0.1±0.003 | 0.3±0.006 |
| MC-GR | 48.5±0.046 | 44.0±0.046 | 1.0±0.009 | 0.7±0.007 | 1.1±0.010 | 1.1±0.010 | 2.6±0.015 | 0.1±0.003 | 0.3±0.005 |
| $M_5B_3$ | 0.2±0.003 | 0.0 | 0.0 | 0.0 | 2.1±0.009 | 2.4±0.009 | 58.0±0.030 | 36.5±0.029 | 0.0 |
| $M_5B_3$ | 0.2±0.002 | 0.0 | 0.0 | 0.0 | 6.2±0.006 | 4.3±0.005 | 51.8±0.013 | 35.7±0.013 | 0.0 |

GB: Grain Boundary, GR: Grain Interior

Figure 4a shows an APT reconstruction containing the γ matrix, an intergranular $M_5B_3$ boride with two interfaces with γ' precipitates. A composition profile across the grain boundary interface $M_5B_3$/γ', along the arrow denoted #1 in Figure 4a, revealed segregation of Cr, B and Mo to the grain boundary, as shown in Figures 4b and 4c. In particular, Cr exhibits segregation up to 10.0 at.%, whereas the B content at the grain boundary reached approx. 5.0 at.% and Mo only 2.0 at.%. Concurrently, Ni and Al were depleted from the grain boundary as shown in Figure 4b.

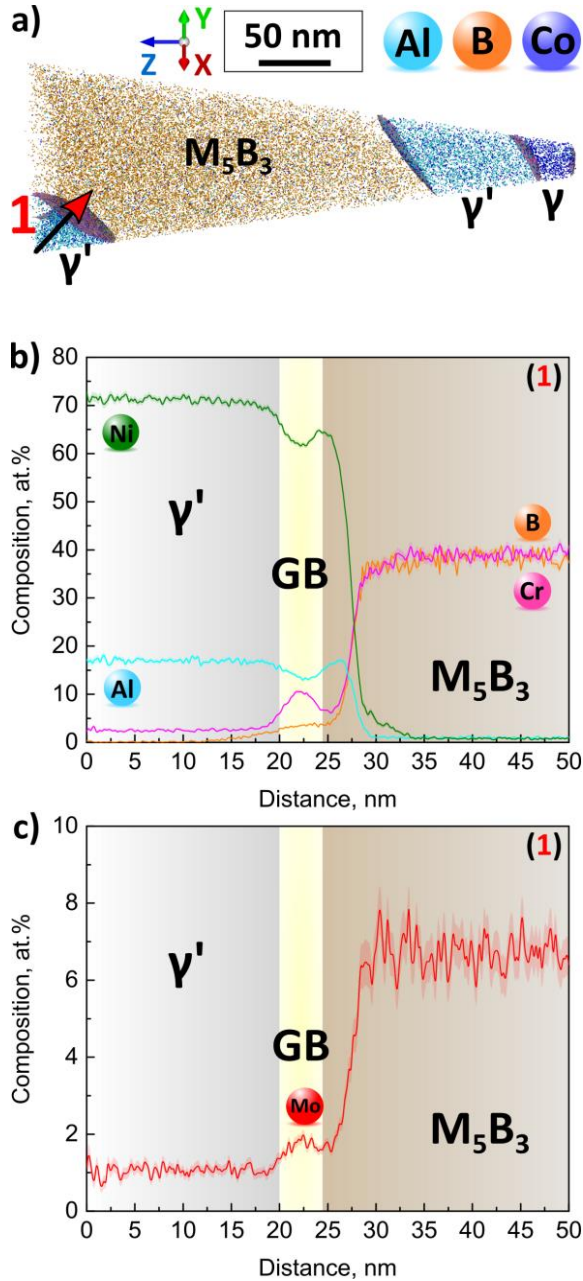

Figure 4: a) APT reconstruction from a grain boundary of STAL15-CC after full heat treatment, showing intergranular $M_5B_3$, γ' and γ with an isosurface at 26.0 at.% Cr. b) and c) Composition profiles across the $M_5B_3$/γ' grain boundary interface corresponding to arrow #1 in Figure 4a. Error bars are shown as lines filled with colour and correspond to the 2σ counting error.

## 3.2. Grain boundary microstructural evolution after creep

Figure 5a shows a backscattered micrograph from a grain boundary of STAL15-CC after creep at 850 °C and 235MPa load. The bright contrast of the MC carbides along the grain boundaries is similarly visible as in the heat-treated condition. However, during creep the MC carbides have decomposed and now they are enveloped by new microstructural layers, exhibiting an intermediate contrast in backscattered electron mode. These new layers are brighter than the γ/γ' microstructure and darker than the MC carbides as clearly shown in Figure 5b, suggesting a compositional difference. These layers are always observed to form between the MC carbides and the intergranular γ' layers. The newly formed layers are more commonly found along the grain boundaries. By contrast, the majority of the intragranular MC carbides was enveloped only within γ' layers after the creep deformation, as shown in Figure 5c.

These layers with intermediate contrast in backscattered electron mode correspond to the brittle TCP η phase as confirmed by EBSD and TEM. From the EBSD analysis, the η phase was found to exhibit a hexagonal crystal structure similar to that of $Ni_3Ti$. Further careful examination of the η layers in the SEM has also shown that the intergranular η layers do not always have a single orientation but they exhibit different orientations with respect to the two adjacent grains as shown from the backscattered micrograph in Figure 5b. This observation was also confirmed by EBSD analysis as shown by the inversed pole figure (IPF) map in Figure 6a and the corresponding phase map in Figure 6b. It is apparent that their formation is strongly related to the decomposition of intergranular MC carbides at 850 °C. The decomposition of the MC carbides led also to the reduction of their size, resulting in carbides with sizes below 500 nm compared to the few microns-sized carbides observed in the heat-treated condition.

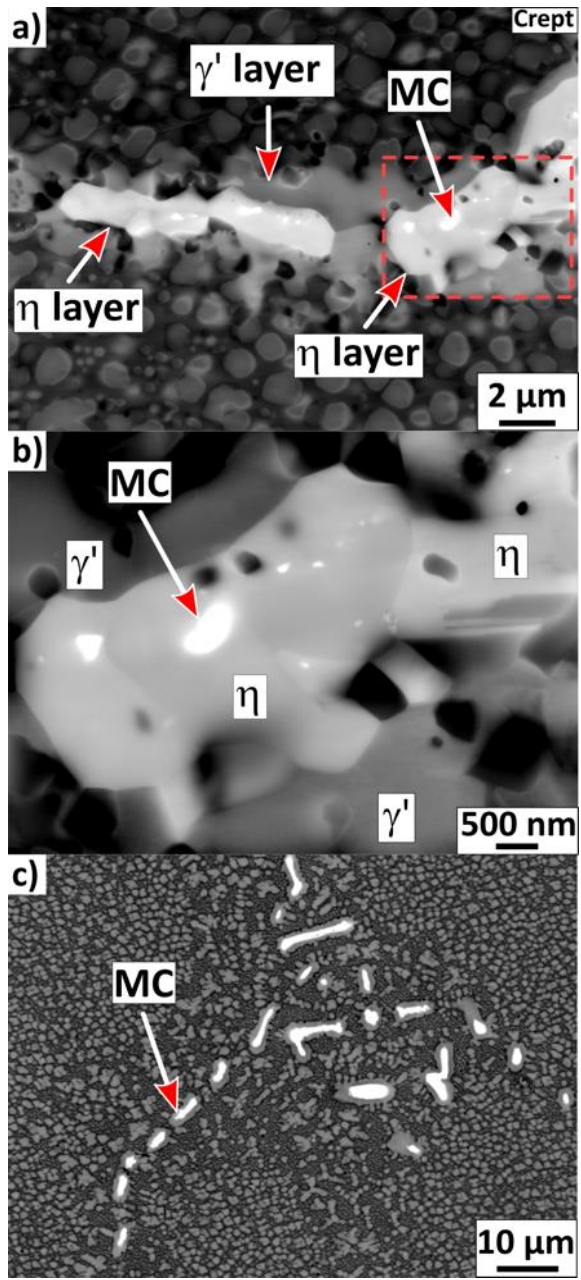

Figure 5: Backscattered electron micrographs of STAL15-CC alloy after creep at 850 °C. a) η layers forming at grain boundaries in the vicinity of decomposed MC carbides. b) Detail of the η layer denoted with a dashed red box in Figure 5a, showing grain boundaries within the η layer. c) Intragranular MC carbides enveloped within γ' layers.
10

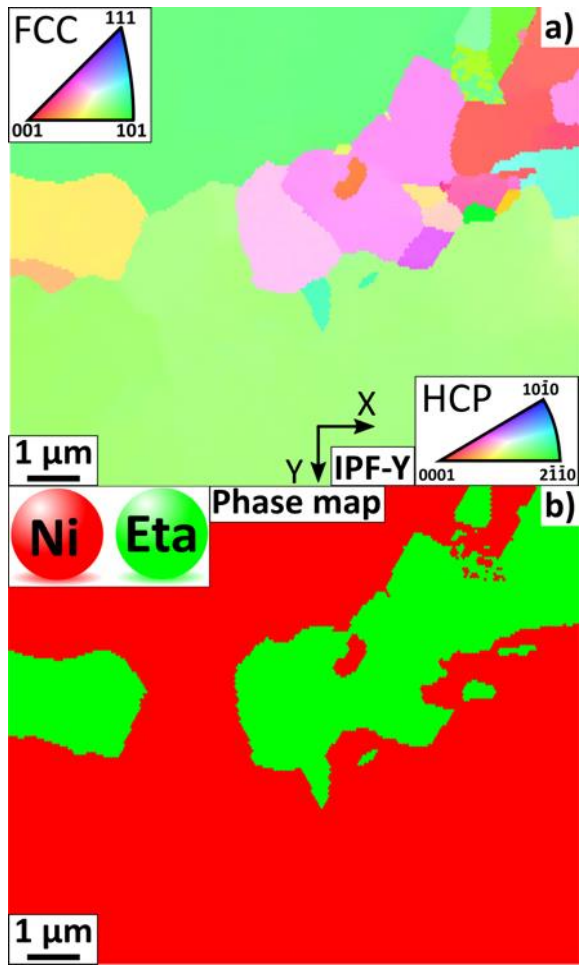

Figure 6: EBSD inverse pole figure (IPF) map (reference vector in vertical Y direction) alongside phase map confirming the various orientations of the η layers along the grain boundaries. 'Ni' refers to the Ni base solid solution matrix.

TEM analysis from a grain boundary that was decorated with similar intergranular phases as those in Figure 5a, evidenced that the η layers exhibit a $D0_{24}$ hexagonal crystal structure, as shown in Figure 7. For comparison, diffraction patterns of the intergranular MC carbides and the γ' layers are provided in Figure 7. The TEM analysis has also shown the formation of intergranular $M_{23}C_6$ carbides within parts of this complex microstructure. $M_{23}C_6$ carbides were not observed in the heat-treated condition, but only after creep deformation as a result of the decomposition of MC carbides.

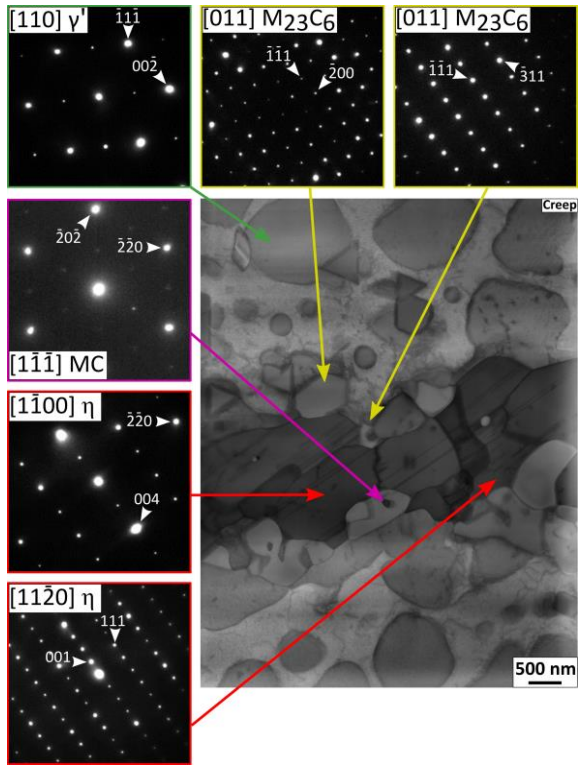

Figure 7: Transmission electron micrograph from a grain boundary after creep at 850 °C with corresponding selected area diffraction patterns of MC, η, $M_{23}C_6$ and γ' precipitates.

By contrast, $M_5B_3$ borides were found to be microstructurally stable, compared to MC carbides. They were seen to coarsen during creep deformation as shown in Figure 8a. However, less often a γ' depleted region was observed in the vicinity of coarsened borides after creep, as shown in Figure 8b. Such γ' depleted region can potentially act as crack initiation sites during creep, but in this study no cracks were observed in the vicinity of γ' depleted zones.

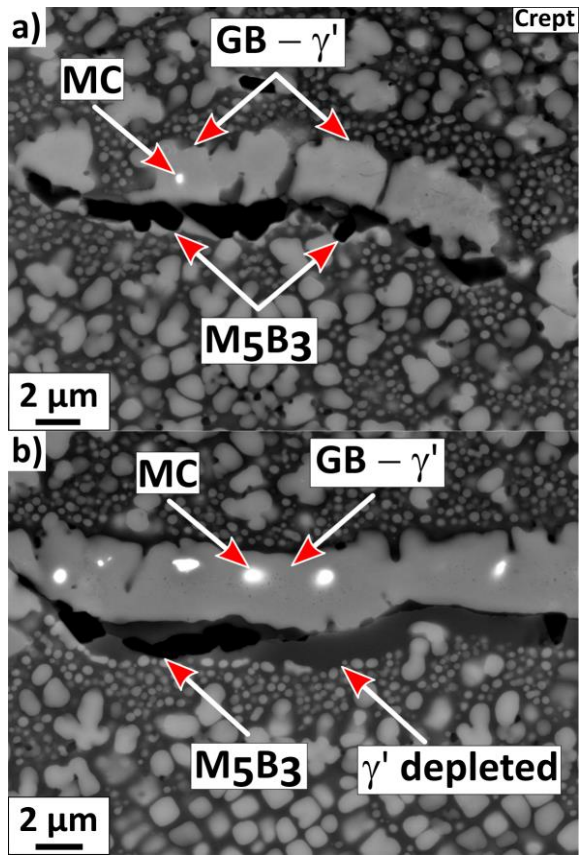

Figure 8: Backscattered electron micrographs from grain boundaries after creep at 850 °C showing a) coarsened $M_5B_3$ borides along the grain boundary and b) γ' depleted regions in the vicinity of coarsened borides.

### 3.3. Compositional evolution of after creep

Several APT specimens specifically prepared in these regions were analysed for each phase in order to investigate the compositional evolution of the different phases after creep under 235MPa load at 850 °C. First, we observed a substantial compositional difference for the intergranular MC carbides, denoted as MC-GB in Table 2. After decomposition of the MC carbides, the Ta content was reduced from approximately 44.0 at.% in the heat-treated condition to approximately 36.0 at.% after creep. Interestingly, the Hf and Zr content increased in the intergranular MC carbides after creep exposure. In particular, Hf was increased from 1.5 at.% to a content varying between 5.0 and 6.1 at.%, as shown in Table 2. In addition, Zr increased from approximately 1.0 at.% to an average 5.5 at.%. Elements such as W, Mo and Cr did not exhibit any dramatic variation and their content remained similar to that at the heat treatment condition.

Comparison of the composition of the MC carbides after creep deformation and between intergranular and intragranular (denoted as MC-GR) revealed that the intragranular carbides were not affected and maintained their initial composition. This supports our findings that the η phase formed primarily at the grain boundaries and not within the grains. Only in very few cases, η layers were observed around MC carbides within the grains as shown in Figure 9. In the case of $M_5B_3$, APT analyses have also shown a compositional evolution, although not as substantial as in the case of MC carbides. Initially, the W content varied between 2.1 and 6.2 at.% and Mo was between 2.4 and 4.3 at.%. After creep, the content for both elements decreased approximately below 1.0 at.%. Interestingly, the Cr content increased to about 65.0 at.% from initially varying between 51.8 and 58.0 at.%. The composition of the $M_5B_3$ borides is shown in detail in Table 2.

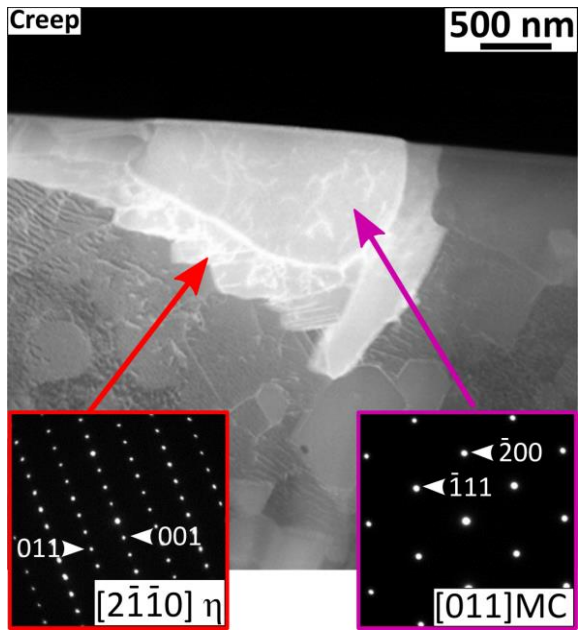

Figure 9: Transmission electron micrograph from an MC carbide within a grain after creep at 850 °C with corresponding selected area diffraction patterns of MC and η.

Table 2: Summary of composition of MC carbides after creep test at 850 °C from grain boundaries (GB) and within grains (GR) and intergranular $M_5B_3$ borides as collected by APT (at.%).

|        | C | Ta | Hf | Zr | W | Mo | Cr | B | N |
|---|---|---|---|---|---|---|---|---|---|
| MC-GB | 49.2±0.024 | 36.6±0.024 | 5.0±0.011 | 5.0±0.011 | 0.8±0.004 | 1.3±0.006 | 1.0±0.005 | 0.4±0.003 | 0.3±0.003 |
| MC-GB | 48.6±0.028 | 35.6±0.027 | 6.1±0.013 | 5.8±0.013 | 0.7±0.005 | 1.2±0.006 | 0.8±0.005 | 0.4±0.004 | 0.2±0.003 |
| MC-GB | 48.7±0.031 | 35.8±0.030 | 6.1±0.015 | 5.9±0.015 | 0.6±0.005 | 1.2±0.007 | 0.7±0.005 | 0.4±0.004 | 0.2±0.003 |
| MC-GR | 47.7±0.030 | 44.6±0.030 | 0.4±0.004 | 0.7±0.005 | 1.2±0.007 | 1.4±0.007 | 3.5±0.011 | 0.0 | 0.2±0.002 |
| MC-GR | 47.3±0.030 | 44.4±0.030 | 0.4±0.004 | 0.6±0.005 | 1.6±0.007 | 1.4±0.007 | 3.7±0.011 | 0.1±0.001 | 0.2±0.002 |
| $M_5B_3$ | 0.1±0.001 | 0.0 | 0.0 | 0.0 | 0.8±0.004 | 0.6±0.003 | 64.9±0.019 | 32.9±0.019 | 0.0 |
| $M_5B_3$ | 0.1±0.002 | 0.0 | 0.0 | 0.0 | 1.1±0.009 | 0.9±0.008 | 65.4±0.040 | 32.0±0.039 | 0.0 |

GB: Grain Boundary, GR: Grain Interior

APT analysis was also performed on the intergranular η layers and their composition from three different locations is given in Table 3. For comparison, the composition from an intergranular γ' layer is also shown in Table 3. It can be seen that Ta substituted Al resulting in the formation of η layers, with the Ta content reaching up to approximately 14.0 at. % in the η phase, compared to 5.3 at.% in the γ' layer. The Al content decreased from 15.6 at.% in the γ' layer to approximately 9.5 at.% in the η phase. Besides, Co was found to increase in the η phase, whereas no substantial fluctuation in the content of other elements such Cr, W, Hf and Mo was observed.

Table 3: Summary of composition of intergranular η phase and γ′ after creep test at 850 °C as collected by APT (at.%).

| | Ni | Ta | Al | Co | Cr | W | Hf | Mo | Zr |
|---|---|---|---|---|---|---|---|---|---|
| η | 70.7±0.019 | 13.2±0.014 | 10.2±0.012 | 3.7±0.008 | 0.7±0.003 | 0.5±0.003 | 0.3±0.002 | 0.1±0.001 | 0.1±0.001 |
| η | 70.4±0.012 | 13.4±0.009 | 9.7±0.007 | 4.0±0.005 | 0.9±0.002 | 0.5±0.002 | 0.4±0.002 | 0.1±0.001 | 0.1±0.001 |
| η | 70.1±0.053 | 14.1±0.041 | 9.0±0.033 | 4.4±0.024 | 1.1±0.012 | 0.3±0.007 | 0.2±0.006 | 0.1±0.003 | 0.1±0.003 |
| γ′ creep-GB | 73.6±0.081 | 5.3±0.041 | 15.6±0.067 | 2.1±0.027 | 1.6±0.023 | 1.1±0.019 | 0.1±0.004 | 0.4±0.011 | 0.0 |

GB: Grain Boundary

Finally, APT analyses were conducted on the $M_{23}C_6$ carbides which have been newly formed during creep. They were found to be mainly Cr-rich, while Ni, Mo, W and Co were also found but in much lower concentration as shown in Table 4 for two different $M_{23}C_6$ carbides. Also, minor amounts of B were detected within the $M_{23}C_6$ carbides, similar to previous observations reported for such types of carbides [17]. An APT reconstruction containing an $M_{23}C_6$ carbide alongside γ matrix and γ' precipitate interfaces is illustrated in Figure 10. A composition profile across the $M_{23}C_6$/γ did not reveal any particular enrichment of any element at this specific interface.

Table 4: Summary of composition of $M_{23}C_6$ carbides formed after creep test at 850 °C as collected by APT (at.%).

|  | Cr | C | B | Ni | Mo | W | Co |
|---|---|---|---|---|---|---|---|
| $M_{23}C_6$ | 71.5±0.030 | 18.2±0.030 | 1.6±0.002 | 3.3±0.006 | 2.6±0.005 | 2.0±0.004 | 0.8±0.002 |
| $M_{23}C_6$ | 68.9±0.015 | 19.8±0.014 | 1.9±0.001 | 3.3±0.003 | 2.5±0.002 | 2.8±0.003 | 0.8±0.001 |

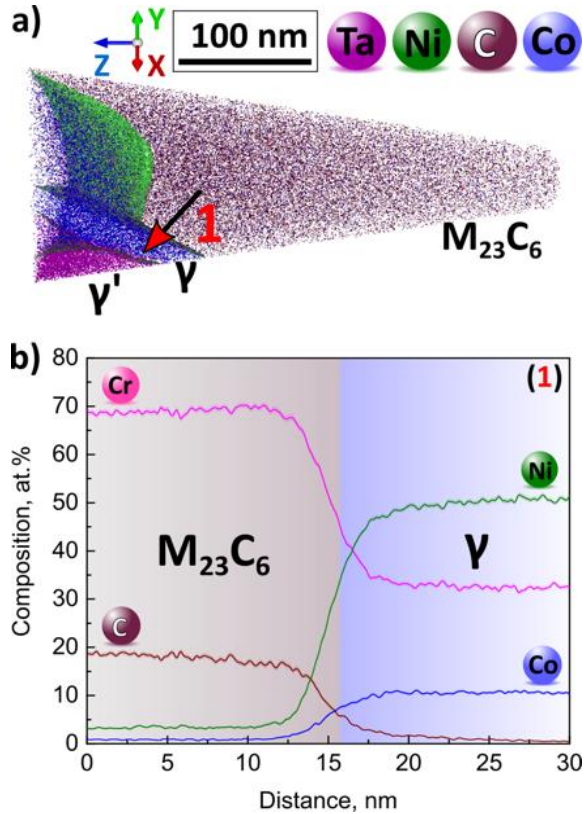

Figure 10: APT reconstruction from a grain boundary of STAL15-CC after creep, showing intergranular $M_{23}C_6$, γ' and γ with isosurfaces at 12.0 and 75.0 at.% Ni. b) Composition profile across the $M_{23}C_6$/γ interface corresponding to arrow #1 in Figure 9a. Error bars are shown as lines filled with colour and correspond to the 2σ counting error.

## 4. Discussion

### 4.1. Stability and decomposition of secondary intergranular phases

It becomes apparent that the decomposition of the MC carbides when resulting in the formation of brittle η layers along the grain boundaries over prolonged exposure times becomes a critical design parameter, which affects the creep life of polycrystalline superalloys. It is necessary for MC carbides to remain stable, and this can be achieved by better controlling and estimating their compositional evolution over time under load at high temperature. It is well known that elements such as Ta, Ti, Hf, W, V or Nb combine with carbon and form MC carbides at high temperatures directly from the liquid phase [21]. Their final composition is often reported to be complex and more than one of the abovementioned elements participates in their formation, as shown also in this study. As it is expected, the final composition of the carbide will be responsible for its stability. The order from the most to the least thermodynamically stable MC carbide is known to be as TaC, NbC, TiC, VC [42].

In this study, we found a reduction of the Ta content in the MC carbide during its decomposition and an increase of the Hf and Zr content as clearly revealed in Figure 11. This observation implies that Hf and Zr-rich carbides are more stable than carbides that are only rich in Ta at 850 °C, listing Hf as the most desirable element for the formation of MC carbides, in terms of stability. Besides, additions of Hf exerts a beneficial effect on the shape and lattice parameter of MC carbides. In particular, the lattice misfit at the MC/matrix interface increases as the Hf content in MC carbides increases leading to blocky MC carbides [43]. Blocky carbides are known to be less detrimental to the mechanical performance of superalloys compared to script-like carbides [44,45]. Recently, it was shown that oxidized MC carbides act as crack initiation sites in polycrystalline superalloys [23–25]. Hf-rich MC carbides are expected to be more difficult to oxidize than Ta-rich carbides under given temperature. Recently, Zhang et al. [46] experimentally confirmed that the oxidation process of TaC initiates at 750 °C whereas HfC oxidizes at 800 °C. The weight gain of HfC increases gradually compared to a very sharp increase in the case of TaC. In the same study, it was shown that a $(Ta_{50}Hf_{50})C$ compound has the best oxidation resistance among these three carbides and its oxidation initiated at 940 °C. During oxidation of TaC to $Ta_2O_5$, a volume expansion of a factor of 2.5 takes place whereas for the HfC to $HfO_2$ the volume expansion is of a factor 1.7 only [47]. The higher volume mismatch in the case of $Ta_2O_5$ compared to $HfO_2$ is less desirable for polycrystalline superalloys, since larger strain gradients will form, hence facilitating crack initiation [24,25]. Also, weight gain experiments revealed that TaC-based composites are heavily oxidized compared to HfC-based ones [48]. Although a pure Hf-rich MC carbide is practically unlikely to form due to the high amounts of Hf required in the bulk composition, our results strongly indicate that MC carbides with a Hf content higher than achieved in this study to be beneficial to the mechanical performance of superalloys.

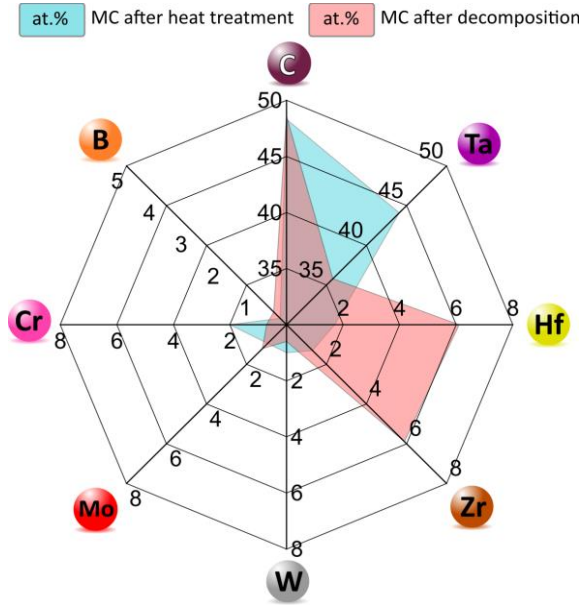

Figure 11: Radar-plot showing the compositional evolution of intergranular MC carbides after their decomposition at 850 °C during creep.

Similar to the behaviour of Hf, the Zr content was also found to increase in MC carbides after creep. Zr is known to be beneficial to the mechanical properties of superalloys by increasing grain boundary cohesion and by impeding the undesirable effect of oxygen at elevated temperatures [49–51]. Our observations also suggest that another beneficial effect of Zr lies in its participation in the formation of MC carbides. Through this effect Zr promotes and facilitates formation of stable MC carbides over prolonged exposure service times.

In contrast, the $M_5B_3$ borides were found to be more stable compared to MC carbides and no signs of decomposition and formation of a new phase were observed after nearly 3000 h under creep at 850 °C. These findings support the more desirable presence of borides along the grain boundaries, compared to MC carbides. However, the fact that γ' depleted zones were found in some limited cases needs to be taken into account. Borides are Cr-rich and were shown to coarsen over time. This process will result in high amounts of Cr to be consumed from the surrounding matrix, resulting in a Cr depleted region. As a consequence, the Ni solubility will locally increase in this region leading to dissolution of γ' precipitates as shown in Figure 8b. Similar observations were made for the coarsening process of $M_{23}C_6$ carbides at grain boundaries resulting in γ' depleted zones [52]. In our study, no cracks were found to form in such γ' depleted zones, but the possibility of those forming at longer exposure times cannot be dismissed. Thus, it is critical to carefully control the volume fraction of the above secondary phases in the heat-treated state. It should be noted that in order to fully rationalize the observed precipitation evolution, additional microstructural observations are necessary for the alloy in the absence of any external applied load, i.e. static thermal exposure at 850 °C for 3000h. In this way, the effect of the plastic deformation and applied stress on the phase evolution can also be studied.

### 4.2. Formation of η phase and its role on creep performance

In order to understand the formation of the intergranular η layers, it is first important to carefully observe the grain boundary character prior to creep. Figure 1a shows intergranular MC carbides enveloped within γ' layers. As the MC carbide decomposes, the Ta content increases locally, however, diffusion of Ta is limited by the presence of the γ' layer. As a consequence, the Ta piles up and at the interface between the MC and γ', the solubility limit of Ta is reached, which leads to the formation of η phase as shown in Figure 5a. At the same time, C that is released by the decomposition of the MC carbide can diffuse more easily through the γ' layer and further along the grain boundary. It combines then with the available Cr at the grain boundary interfaces, as shown in Figure 4b, and forms the Cr-rich $M_{23}C_6$ as confirmed by TEM and APT analysis.

A higher volume fraction of $M_{23}C_6$ carbides compared to what it was found by TEM should be expected after the deformation. In Figure 5a and 5b, many particles with a dark contrast can be observed around the η layers, which could potentially turn out to be $M_{23}C_6$. Yet, SEM observations alone do not allow for unambiguously distinguishing between the Cr-rich $M_{23}C_6$ carbides and $M_5B_3$ borides, both appearing with a similarly dark contrast in backscatter imaging. However, the shape and size of these dark particles in Figures 5a and 5b are similar to those identified by TEM shown in Figure 7. It is hence likely that these particles are also $M_{23}C_6$ carbides. No C was detected in the γ' or the η layers. Some C can potentially reside at the grain boundaries of the recrystallized-like η layers. However, we were not able to capture this information from within the APT analyses reported here. However, the majority of C, if not all, was likely consumed during the formation of $M_{23}C_6$ carbides.

It is apparent that the presence of the intergranular γ' layers can trigger formation of η layers during the decomposition of carbides. These layers are believed to be beneficial to the mechanical performance of superalloys. In particular, when brittle MC carbides are enveloped within γ' layers, creep ductility and fracture toughness are improved [27]. It was also shown that such layers are desirable because they retard crack initiation at grain boundaries by preventing accumulation of plastic deformation at critical interfaces such as MC/γ'/γ [39]. However, this effect only applies up to temperatures where MC decomposition does not take place or proceeds very slowly. In this particular study, it was shown that decomposition of carbides takes place at 850 °C and cracks forming at intergranular η layers as shown in Figure 12. This observation points towards the design direction, where the presence of γ' layers appears to be undesirable for polycrystalline superalloys that are required to maintain their microstructure at 850 °C or higher temperatures.

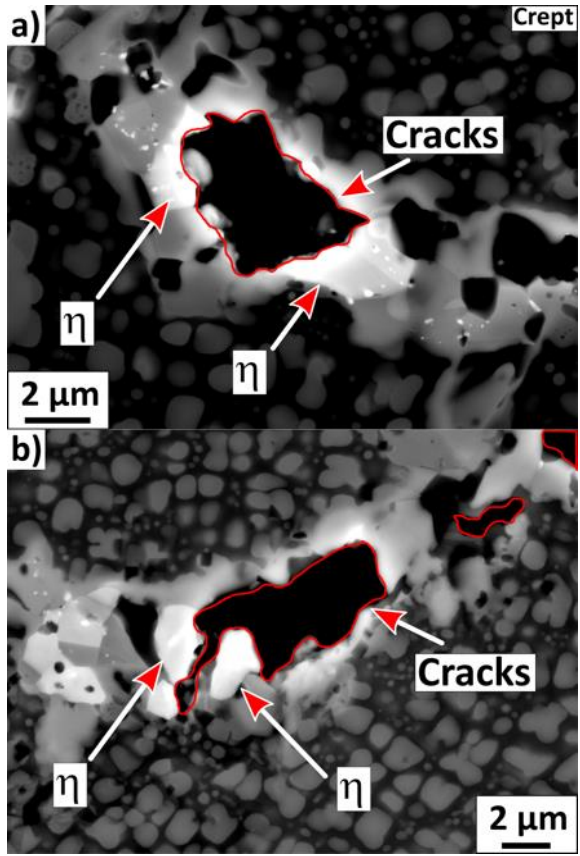

Figure 12: Backscattered electron micrographs of STAL15-CC alloy after creep at 850 °C showing secondary cracks forming at grain boundaries in which η layers have formed.

Although the η phase forming as a result of the decomposition of MC carbides has been reported before [53–56], we present here new insights about its coalescence at the grain boundaries. In Figures 5b and 6a, it was shown that the η layers exhibit various orientations with regard to the adjacent two neighbour grains. As a consequence, these intergranular η layers can act as recrystallised brittle grains and deteriorate the mechanical performance of polycrystalline superalloys. Cracks were mainly found to initiate at grain boundaries during creep. Due to oxidation taking place at the main crack, it is not possible to evaluate completely the effect of the η layers on the creep performance. However, information about the detrimental role of the η layers on crack formation could be extracted from secondary cracks. In many cases, cracks were initiated at the intergranular η layers, most likely due to low cohesion between η and γ' as shown in Figure 12. In particular, secondary cracks were shown to initiate at grain boundaries with η layers being present. Until now, deterioration of the mechanical performance of crept polycrystalline superalloys was only attributed to the presence and brittle nature of this TCP phase. Our observations of the recrystallisation-like regions along grain boundaries improve our understanding on how and why η layers are acting on damage initiation in polycrystalline superalloys. In addition, our study provides new insights on the compositional evolution during deformation at elevated temperatures of secondary phases such as borides and carbides, which can exert a significant influence over the overall properties of polycrystalline superalloys. These observations should be used to guide materials modelling approaches to further improve life-predictions of safety-critical components and improve our understanding of these high performance materials.

**Conclusions**

We have studied the microstructural evolution of the grain boundary composition and precipitation states and their effects on damage initiation in the polycrystalline nickel-based superalloy STAL15-CC during creep at 850 °C. The compositional decoration and grain boundary phases were investigated by TEM and APT and compared with those after creep fracture at 850 °C. The following conclusions were drawn:

- Primary intergranular Ta-rich MC carbides have decomposed after approximately 3000 hours at 850 °C. Their chemistry was altered substantially from the heat treated condition resulting in the Ta content to drop to approximately 8.5 at.% in average. Hf was increased from 1.5 at.% to a content varying between 5.0 and 6.1 at.% and Zr was increased from approximately 1.0 at.% to an average 5.5 at.%.
- The decomposition of the primary MC carbide led to the formation of the brittle TCP η phase with a structure consistent with the $D0_{24}$ hexagonal structure as confirmed by TEM and $Ni_3Ta$ chemistry revealed by APT.
- EBSD analysis has shown that intergranular η phase layers are forming with various misorientations with regard to the adjacent two neighbour grains acting as brittle recrystallized zones at the grain boundaries and thus, as crack initiation site under creep conditions.
- The presence of the γ′ layers at the grain boundaries of the fully heat treated microstructure is rather detrimental to the long-term mechanical properties and they are responsible for the formation of the intergranular η layers.

- The second product of the decomposition of primary MC carbide was the formation of intergranular Cr-rich $M_{23}C_6$ carbides. No $M_{23}C_6$ carbides were found at the fully heat treated microstructure.
- Primary intergranular Cr-rich $M_5B_3$ borides were not dissolved after creep deformation at 850 °C for 3000 h, instead they have coarsened. Their composition was only slightly altered, particularly the Cr content increased after creep. Coarsening of borides resulted in γ′ depleted zones in few cases, however no cracks were formed there.

**Acknowledgements**

P.K. thanks Siemens Industrial Turbomachinery for provision of the material and performing the creep tests. P.K., B.G. and D.R. are deeply grateful to the MPG for the funding of the Laplace project. Uwe Tezins and Andreas Sturm are also acknowledged for their support on APT experiments. Finally, Monika Nellessen and Katja Angenendt are thanked for their support on EBSD experiments.